\documentclass[conference]{IEEEtran}

\usepackage{times}
\usepackage{amsmath}
\usepackage{amssymb}
\usepackage{algorithm}
\usepackage{algorithmic}
\usepackage{graphicx}
\usepackage{subfigure}
\usepackage{verbatim}
\usepackage[usenames]{color}
\usepackage{cite}
\usepackage{url}
\usepackage{soul}
\usepackage{todonotes}

\begin{document}

\title{Range-Free Localization with the Radical Line}

\author{
Hongyang Chen$^{1,2}$, Y. T. Chan$^3$, H. Vincent Poor$^4$, and Kaoru Sezaki$^{1,2}$\\
$^1$Institute of Industrial Science, The University of Tokyo, Tokyo, Japan\\
$^2$CREST, JST, Japan\\
$^3$Department of Electrical and Computer Engineering, Royal Military College of Canada, Kingston, Canada\\
$^4$Department of Electrical Engineering, Princeton University, Princeton, NJ, USA\\

Email: \ hongyang@mcl.iis.u-tokyo.ac.jp, chan-yt@rmc.ca, poor@princeton.edu, sezaki@iis.u-tokyo.ac.jp
}

\maketitle

\begin{abstract}
Due to hardware and computational constraints, wireless sensor networks (WSNs) normally do not take measurements of
time-of-arrival or time-difference-of-arrival for range-based localization.
Instead, WSNs in some applications use range-free localization for simple but less accurate determination
of sensor positions. A well-known algorithm for this purpose is the centroid algorithm. This paper presents a range-free localization technique based on the radical line of intersecting circles. This technique provides greater accuracy than the centroid algorithm, at the expense of a slight increase in computational load. Simulation results show that for the scenarios studied, the radical line method can give an approximately 2 to 30\%
increase in accuracy over the centroid algorithm, depending on whether or not the anchors have identical ranges, and on the value of DOI.

\end{abstract}


%
\IEEEpeerreviewmaketitle

\section{Introduction}

A wireless sensor network (WSN) \cite{apit} typically consists of anchors and sensors communicating with each other. An anchor broadcasts its position coordinates, together with operating instructions, to the sensors. A sensor needs to determine its position to report to the anchors. Position determination can come from time-of-arrival, time-difference-of-arrival or angle-of-arrival measurements \cite{mdtoa}. But when the sensors are low cost, low power and expandable units, with  limited resources for computation, they often rely on range-free (RF) localization instead \cite{GPS-less}.

In RF localization, a sensor $P$ determines its unknown position $P = {[x,y]^T}$ from $N$ in-contact
anchors $a_i$ at known positions $a_i = {[x_i,y_i]^T}$ and having radio ranges $R_i, i=1,...,N$. The sensor position must satisfy
\begin{eqnarray}\label{eq:range}
\begin{gathered}
  {\left\| {P - {a_i}} \right\|^{1/2}} = {[{(x - {x_i})^2} + {(y - {y_i})^2}]^{1/2}} \hfill \\
  \;\;\;\;\;\;\;\;\;\;\;\;\;\;\;~~~ \leqslant {R_i},\;\;\;\;\;i = 1,2, \ldots ,N. \hfill \\
\end{gathered}
\end{eqnarray}
Solving (\ref{eq:range}) requires nonlinear programming, and there is not a unique answer. The Centroid Algorithm (CA) [3] gives a simple estimate $\hat P = {[\hat x,\hat y]^T}$, where
\begin{eqnarray}\label{eq:cae}
\hat x = \frac{{\sum\limits_{i = 1}^N {{x_i}} }}
{N} ~~and ~~ \hat y = \frac{{\sum\limits_{i = 1}^N {{y_i}} }}
{N}.
\end{eqnarray}
But $\hat P$ from (\ref{eq:cae}) sometimes is outside the region of intersections (RI) of the circles centered at $a_i$ with radii $R$, as defined by (\ref{eq:range}). For example, the $\hat P$ in Fig. 1 is outside the RI (shaded area) of the three circles.

This paper proposes a new RF algorithm that has better accuracy than CA, but with a marginal increase in computations. However, the additional computations are well within the capability of present day sensors.

The line joining the intersection points of two circles is the radical line (RL) \cite{RL}. In Fig. 1, the RI contains a segment of the RL of any two circles, and the three RLs meet at a point inside the RI. Indeed, \cite{RL} proves that for three circles whose centers are not collinear, their three RLs always intersect at a point. Although sometimes this point can be outside the RI, it is inside in most cases.

\begin{figure}[t]
\centering
\includegraphics[width=3.5in]{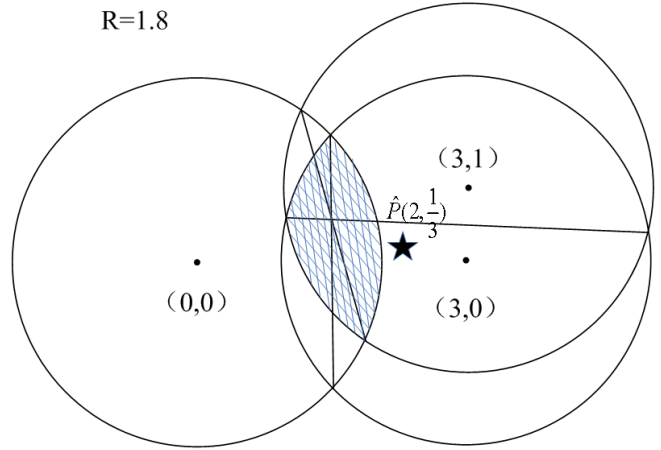}
\caption{The CA and radical line solutions} \label{fig:sensor number and DOI}
\end{figure}

In the following, Section II gives the development of the RL algorithm (RLA). Section III contains simulation results, which show that the RLA is more accurate than the CA, especially when the radii $R_i$ are different. Conclusions are given in Section IV.

\section{The Radical Line Algorithm}
In WSNs, a sensor can determine whether it is in the transmission range of an anchor node according to the beacon signal received from the anchor. Most literature on range-free localization assume a nominal range (or detection range) $R$, i.e., an anchor can communicate with a sensor within $R$ meters from it. However, the actual range in practice is dependent on the propagation conditions. A measure of the variation in range coverage is the degree of irregularity (DOI). Its value denotes the maximum range variation per unit degree change in the direction of radio propagation. Recently, \cite{sinr} gives a condition required for a successful anchor-to-sensor contact. Let $W(a_i)$ be the power received by a sensor from $a_i$, $Q$ be the ambient noise power, and $S$ the interference power in the WSNs. Then there is a contact only if
\begin{eqnarray}\label{sinr}
\frac{{W({a_i})}}
{{Q + S}} \geqslant TH,
\end{eqnarray}
where $TH$ is a hardware dependent threshold.

Let a sensor $P$ be at an unknown position $P = {[x,y]^T}$, in contact with $N$ anchors $a_i$ at known positions $a_i = {[x_i,y_i]^T}$ and having radio ranges $R_i$. Hence $P$ must lie in the RI of the $N$ circles, centered at $a_i$ with radii $R_i$. Depending on $N$, there are three cases to consider.

\subsection{$N > 3$ }
 For $N$ circles, there are $\frac{{N!}}{{2!(N - 2)!}}$ RLs. To reduce computations, RLA selects only the RLs of the two circles whose centers are separated by the largest distance among the $N$ circles. The idea behind this choice
 is that the RL of these two circles will be the shortest, and hence their RL has the highest probability of appearing inside the RI of all the $N$ circles.

Let
\begin{eqnarray}\label{eq:da}
{d_{ij}} = {\left\| {{a_i} - {a_j}} \right\|^{1/2}} = {d_{ji}},\;\;\;i,j = 1, \ldots ,N
\end{eqnarray}
be the distance between the centers of $a_i$ and $a_j$ and let $d_{qk}$ be the maximum of the values in
(\ref{eq:da}). For illustration simplicity, let $q=1$, and $k=2$. Referring to Fig. 2, the end points of the RL are $I_a = {[x_a,y_a]^T}$ and $I_b = {[x_b,y_b]^T}$, and $O = {[x_o,y_o]^T}$ is the intersection between the RL and the line joining $a_1$ and $a_2$.

Let
\begin{eqnarray}\label{eq:do1}
{d_{o1}} = {\left\| {O - {a_1}} \right\|^{1/2}}
\end{eqnarray}
and
\begin{eqnarray}\label{eq:do2}
{d_{o2}} = {\left\| {O - {a_2}} \right\|^{1/2}}.
\end{eqnarray}

It follows that
\begin{eqnarray}\label{eq:xo1}
d_{o1}^2 + {m^2} = R_1^2
\end{eqnarray}
and
\begin{eqnarray}\label{eq:xo2}
d_{o2}^2 + {m^2} = R_2^2.
\end{eqnarray}
Subtracting (\ref{eq:xo2}) from (\ref{eq:xo1}) gives
\begin{eqnarray}\label{eq:xo3}
2(x_2  - x_1 )x_o  + 2(y_2  - y_1 )y_o  = R_1^2  - R_2^2  + k_2  - k_1,
\end{eqnarray}
where
\begin{eqnarray}\label{eq:ki}
k_i  = x_i^2  + y_i^2.
\end{eqnarray}
Further, equating the slopes of the $a_1$ to $O$ and $a_2$ to $a_1$ lines in Fig. 2 yields
\begin{eqnarray}\label{eq:sl1}
\frac{{y_2  - y_o }}{{x_2  - x_o }} = \frac{{y_2  - y_1 }}{{x_2  - x_1 }},
\end{eqnarray}
giving
\begin{eqnarray}\label{eq:sl2}
(y_2  - y{}_1)x_o  - (x_2  - x_1 )y_o  = x_2 (y_2  - y{}_1) - y_2 (x_2  - x_1 ).
\end{eqnarray}
Solving (\ref{eq:xo3}) and (\ref{eq:sl2}) then gives $O(x_o, y_o)$.

Let $d_{12}  = D$. Then
\begin{eqnarray}\label{eq:em}
R_1^2  - d_{o1}^2  = m^2  = R_2^2  - (D - d_{o1} )^2,
\end{eqnarray}

so that
\begin{eqnarray}\label{eq:do2}
d_{o1}  = \frac{{R_1^2  - R_2^2  + D^2 }}{{2D}}
\end{eqnarray}
and
\begin{eqnarray}\label{eq:m2}
m = (R_1^2  - d_{o1}^2 )^{1/2}.
\end{eqnarray}

Now in Fig. 2, the following trigonometric relations hold:
\begin{eqnarray}\label{eq:rel}
\frac{{{x_o} - {x_a}}}
{m} = \frac{{{y_o} - {y_1}}}
{d_{o1}}
\end{eqnarray}
and
\begin{eqnarray}\label{eq:rel2}
\frac{{{y_o} - {y_a}}}
{m} = \frac{{{x_o} - {x_1}}}
{d_{o1}}.
\end{eqnarray}
From (\ref{eq:rel}) and (\ref{eq:rel2}), the coordinates for $I_a$ are
\begin{eqnarray}\label{eq:xa}
{x_a} = {x_o} - \frac{{m}}
{d_{o1}}({y_o} - {y_1})
\end{eqnarray}
and
\begin{eqnarray}\label{eq:ya}
{y_a} = {y_o} + \frac{{m}}
{d_{o1}}({x_o} - {x_1}).
\end{eqnarray}
Following the same procedure gives
\begin{eqnarray}\label{eq:xb}
{x_b} = {x_o} + \frac{{m}}
{d_{o1}}({y_o} - {y_1})
\end{eqnarray}
and
\begin{eqnarray}\label{eq:yb}
{y_b} = {y_o} - \frac{{m}}
{d_{o1}}({x_o} - {x_1}).
\end{eqnarray}

\begin{figure}[t]
\centering
\includegraphics[width=3.5in]{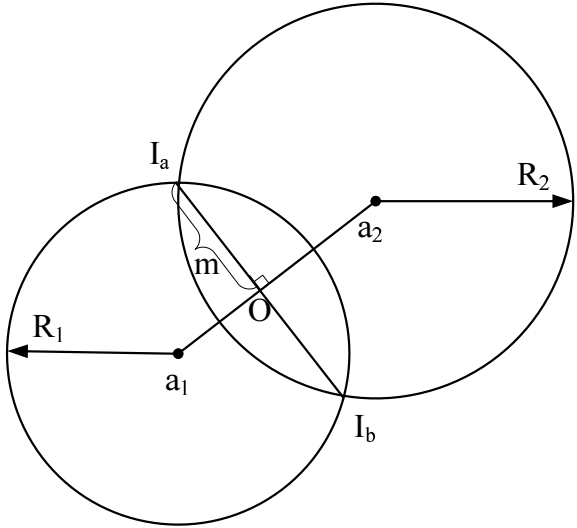}
\caption{The end points of an RL} \label{fig:sensor number and DOI}
\end{figure}

Next, RLA selects $L$ test points ${t_l} = {[{x_l},{y_l}]^T},\;\;l = 1, \ldots ,L,$ on RL, by taking
equal increments between $I_a$ and $I_b$ to give
\begin{eqnarray}\label{eq:xl}
{x_l} = {x_a} + \frac{{l({x_b} - {x_a})}}
{{L + 1}}
\end{eqnarray}
and
\begin{eqnarray}\label{eq:yl}
{y_l} = {y_a} + \frac{{l({y_b} - {y_a})}}
{{L + 1}}.
\end{eqnarray}
 $L$ is a user parameter, depending on the resolution required. In the simulation experiment in Section III,
$L=4$.
At each $t_l$, RLA checks whether $t_l$ is inside the RI, and if not, how far away from the RI it is, by computing the error
\begin{eqnarray}\label{eq:er}
{\varepsilon _{li}} = {\left\| {{t_l} - {a_i}} \right\|^{1/2}} - {R_i} = \left\{ {\begin{array}{*{20}{c}}
   {{\varepsilon _{li}}\;\;\;if\;{\varepsilon _{li}} > 0}  \\
   {0\;\;\;if\;{\varepsilon _{li}} \leqslant 0}  \\
 \end{array} } \right.
 \end {eqnarray}
 and then summing the errors over all $a_i$ to give
 \begin{eqnarray}\label{eq:sumer}
{S_l} = \sum\limits_{i = 1}^N {{\varepsilon _{li}}}.
 \end{eqnarray}
 If an $S_l=0$, the corresponding $t_l$ is inside the RI and is the estimate for $P$. If all $S_l>0$, the RL is not inside the RI of the $N$ circles. It is then necessary to compute the CA errors
  \begin{eqnarray}\label{eq:sumcaer}
{S_c} = \sum\limits_{i = 1}^N {{\varepsilon _{ci}}}
 \end{eqnarray}
 where $\varepsilon _{ci}$ comes from (\ref{eq:er}), with $c = {[\hat x,\hat y]^T}$ from (\ref{eq:cae}) replacing $t_l$. The final estimate for $P$, $\hat P$, comes from choosing the $t_l$ or $c$, whose corresponding $S_l$ or $S_c$ is the minimum.

\subsection{$N=2$ and $N=3$}
When $N=2$, $\hat P$ is the same as $O (x_o, y_o)$.
When $N=3$, RLA computes the intersection of the three RLs. Let that intersection point be $I = {[{x_I},{y_I}]^T}$. Extending Fig. 2 to three circles yields
\begin{eqnarray}\label{eq:3an}
{({x_I} - {x_i})^2} + {({y_I} - {y_i})^2} + {h^2} = {R_i^2},\;\;i = 1,2,3
\end{eqnarray}
where $h^2\leq m^2$.
Subtracting this expression for $i=2,3$ from that for $i=1$ results in
\begin{eqnarray}\label{eq:aib}
AI=b
\end{eqnarray}

where
\begin{eqnarray}\label{eq:x2x3}
A = \left[ {\begin{array}{*{20}{c}}
   {{x_2} - {x_1}} & {{y_2} - {y_1}}  \\
   {{x_3} - {x_1}} & {{y_3} - {y_1}}  \\
 \end{array} } \right],
 \\ and~~
b = \frac{1}
{2}\left[ {\begin{array}{*{20}{c}}
   {{k_2} - {k_1}+R_1^2-R_2^2}  \\
   {{k_3} - {k_1}+R_1^2-R_3^2}  \\
 \end{array} } \right].
\end{eqnarray}

Solving (\ref{eq:aib}) gives
\begin{eqnarray}\label{eq:solv}
I = {A^{ - 1}}b.
\end{eqnarray}
If the determinant of $A$ equals 0, then the three circles are collinear. Or if ${\left\| {I - {a_i}} \right\|^{1/2}} > R_i$ for any $i$, then $I$ is outside the RI. For these two cases, RLA takes the centroid of the two circles with the largest separation as $\hat P$.

\section{   SIMULATION RESULTS}
In the simulation experiments, the WSN has an area of 100 m x 100 m, and contains 100 randomly placed (different for each trial) sensors. For a given number of anchors (NA), occupying random (different for each trial) but known positions, the number of anchors $N$ in contact with an arbitrary sensor can vary from 2 to NA. Some anchors have $R=R_{max}=45m$, and some have $R=0.5R_{max}$. The localization errors decrease with increasing NA. For 100 independent trials, the error as a fraction of $R_{max}$ is
\begin{eqnarray}\label{eq:error}
e(NA) = \frac{1}
{{100}}\sum\limits_{j = 1}^{100} {\left\{ {\frac{{\sum\limits_{i = 1}^{100} {{{\left\| {{p_j}(i) - {{\hat p}_j}(i)} \right\|}^{1/2}}} }}
{100R_{max}}} \right\}}.
\end{eqnarray}
In (\ref{eq:error}), ${p_j}(i)$ is the true \emph{i}th sensor position at trial $j$, and ${{{\hat p}_j}(i)}$
is its estimate.

In an experiment where $DOI=0$, a sensor that lies within the nominal $R_i$ of an anchor is in contact with that anchor. When $DOI\neq0$, the actual $R_i$ is smaller, given by $R_i(DOI)=R_i (1-DOI)$.

Fig. 3 plots $e(NA)$ for both CA and RLA with all anchors' $R=R_{max}$ and as NA varies from 24 to 36, at a $DOI=0$. The results show that RLA has lower $e(NA)$ than CA.

Fig. 4 plots $e(NA)$ for both CA and RLA and all $R=R_{max}$ as NA varies from 24 to 36, at a $DOI=0.1$. The results show that RLA has lower $e(NA)$ than CA.

Fig. 5 plots $e(NA)$ for both CA and RLA with different transmission ranges, i.e., some $a_i$ have $R=R_{max}=45m$ and some have $R=0.5R_{max}$ as NA varies from 24 to 36, at a $DOI=0$. The improvement of RLA over CA is more significant than when all anchors have $R=R_{max}$.

Fig. 6 plots $e(NA)$ for both CA and RLA with different transmission ranges as NA varies from 24 to 36, at a $DOI=0.2$. A comparison of the errors in Figs. 4-6 reveals that RLA has increasing accuracy over CA, when DOI increases. The improvement is more significant when the anchors have different ranges.

\begin{figure}[t]
\centering
\includegraphics[width=3.5in]{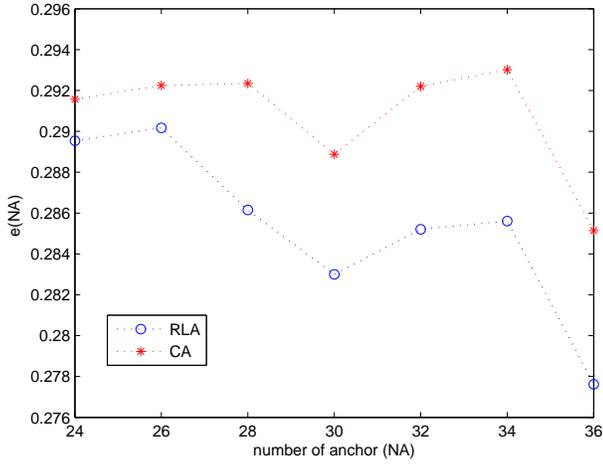}
\caption{The average localization error vs. the number of anchors (DOI=0 and the same transmission range)} \label{fig:sensor number and DOI}
\end{figure}

\begin{figure}[t]
\centering
\includegraphics[width=3.5in]{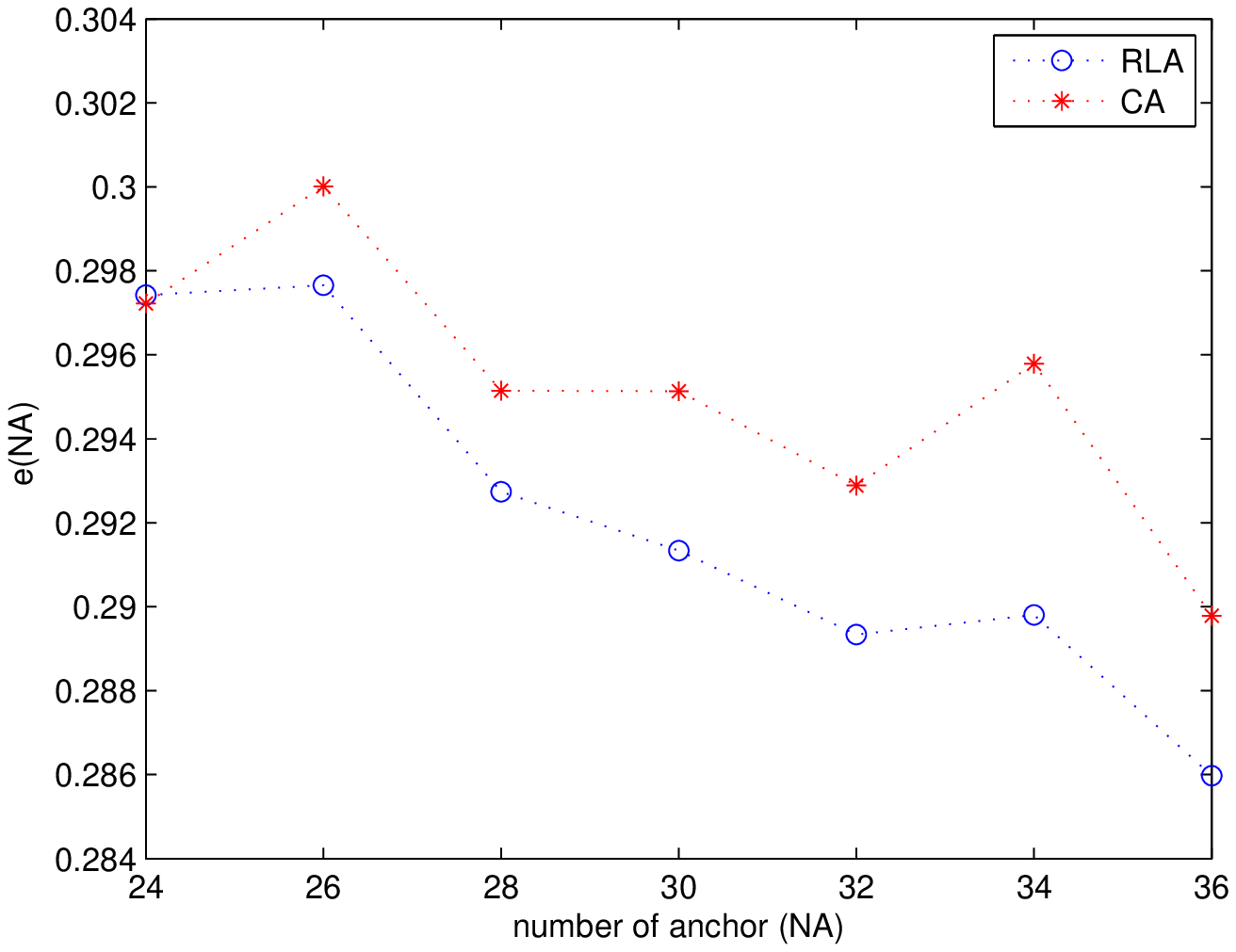}
\caption{The average localization error vs. the number of anchors (DOI=0.1 and the same transmission range)} \label{fig:sensor number and DOI}
\end{figure}

\begin{figure}[t]
\centering
\includegraphics[width=3.5in]{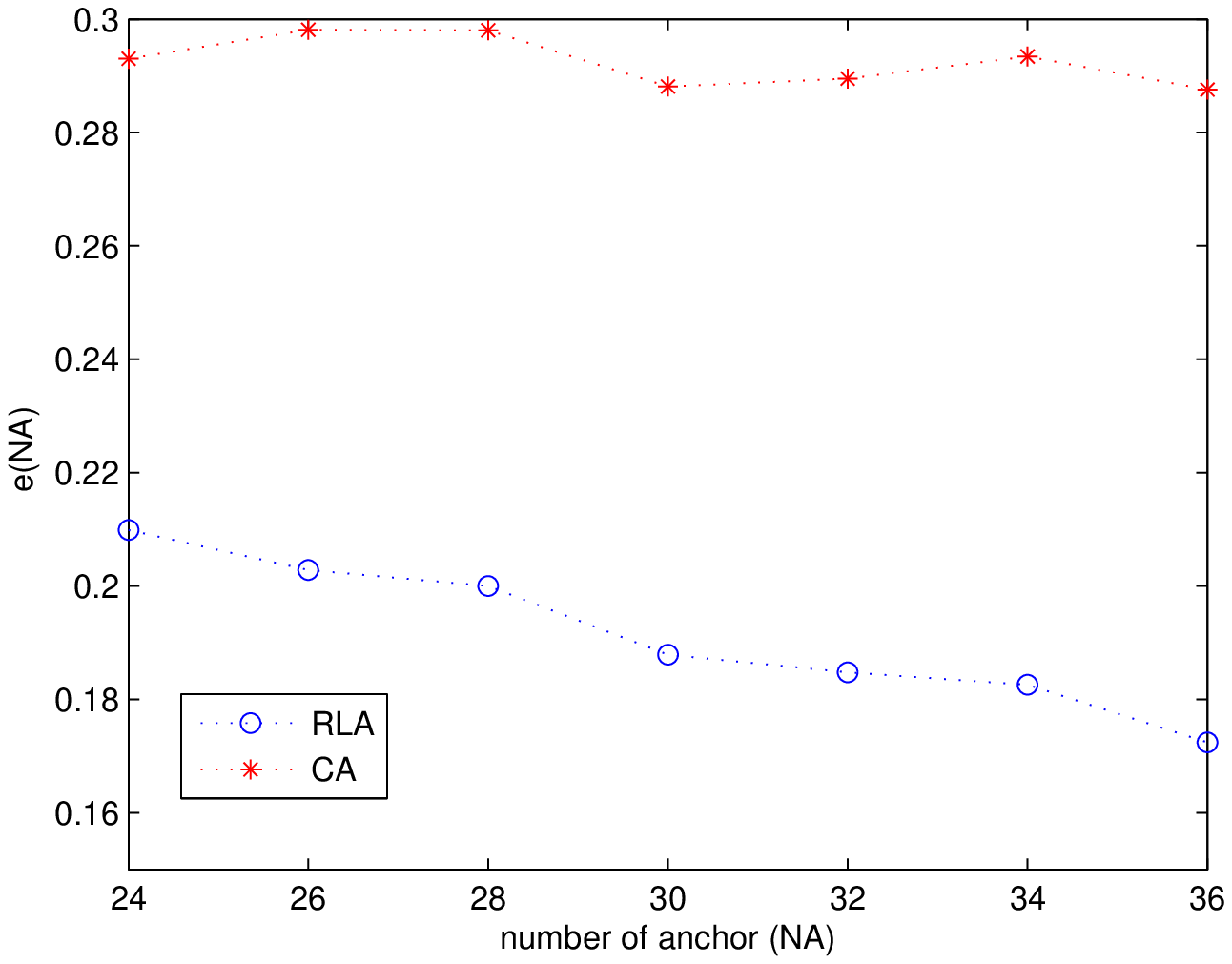}
\caption{The average localization error vs. the number of anchors (DOI=0 and different transmission ranges)} \label{fig:sensor number and DOI}
\end{figure}

\begin{figure}[t]
\centering
\includegraphics[width=3.5in]{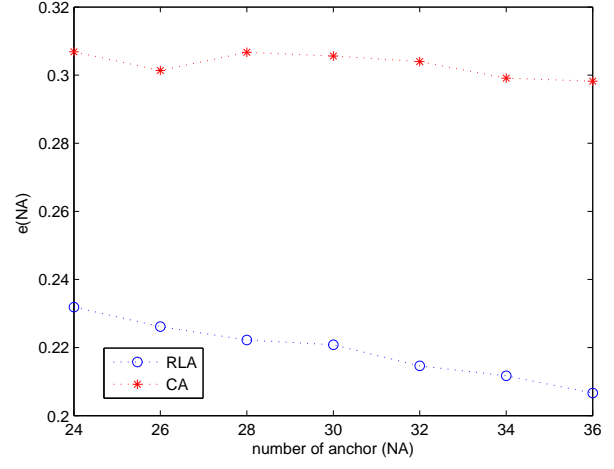}
\caption{The average localization error vs. the number of anchors (DOI=0.2 and different transmission ranges)} \label{fig:sensor number and DOI}
\end{figure}


Fig. 7 is a snapshot of  one trial in the anchor-sensor geometry with $NA=30$ and different transmission ranges, together with the placement of $\hat P$. A dotted line joins $P$ to $\hat P$. Comparing Fig. 7(a) to Fig. 7(b), the dotted lines for RLA are generally
shorter than those for CA.


\begin{figure}
\centering \subfigure[Localization error of RLA (DOI=0.1, $\textrm{error}=0.1929$)]{
\label{fig:subfig:MDS_bad} 
\includegraphics[width=0.53\textwidth]{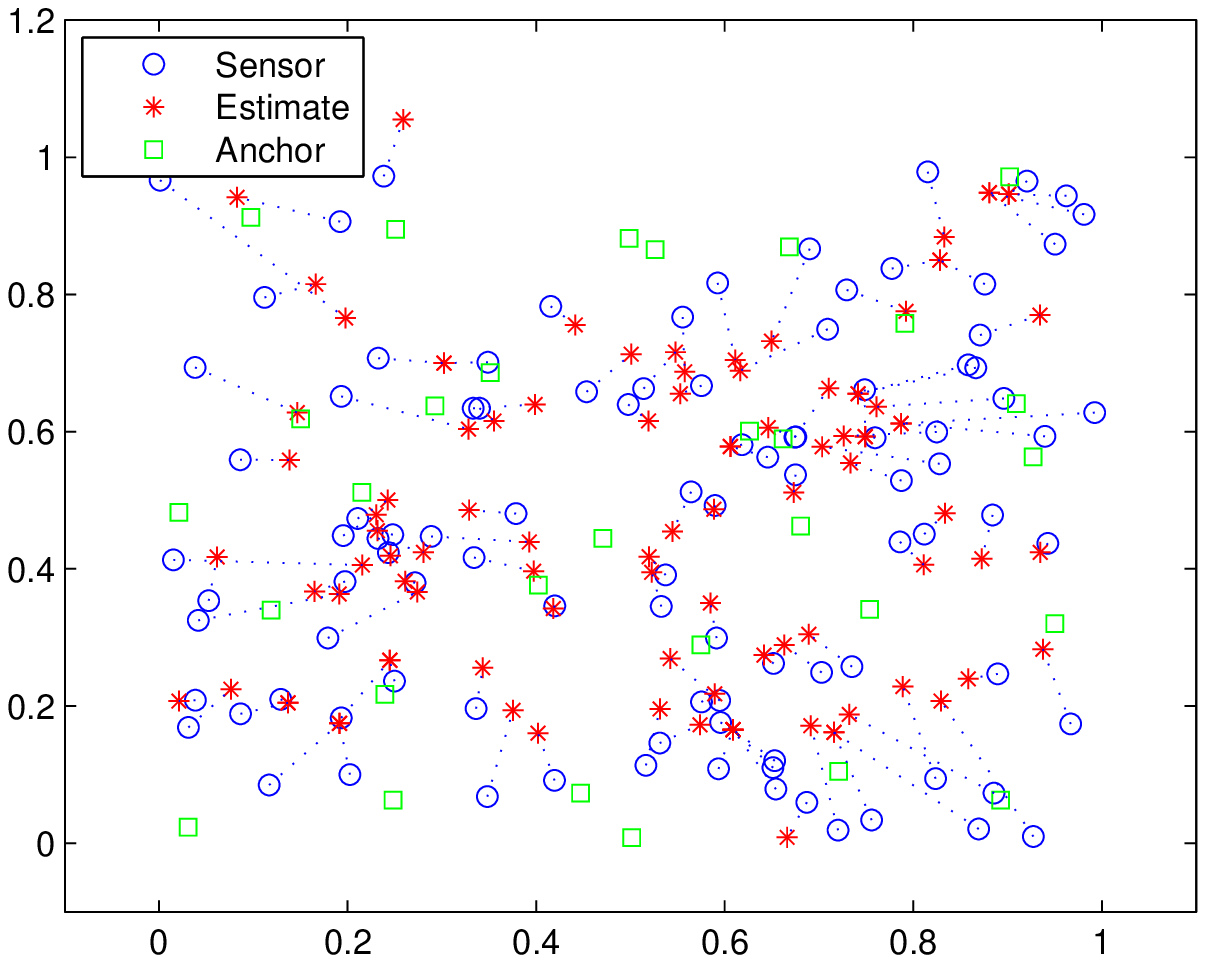}}
\hspace{0in} \subfigure[Localization error of CA (DOI=0.1, $\textrm{error}=0.2872$)]{
\label{fig:subfig:SDP_bad} 
\includegraphics[width=0.53\textwidth]{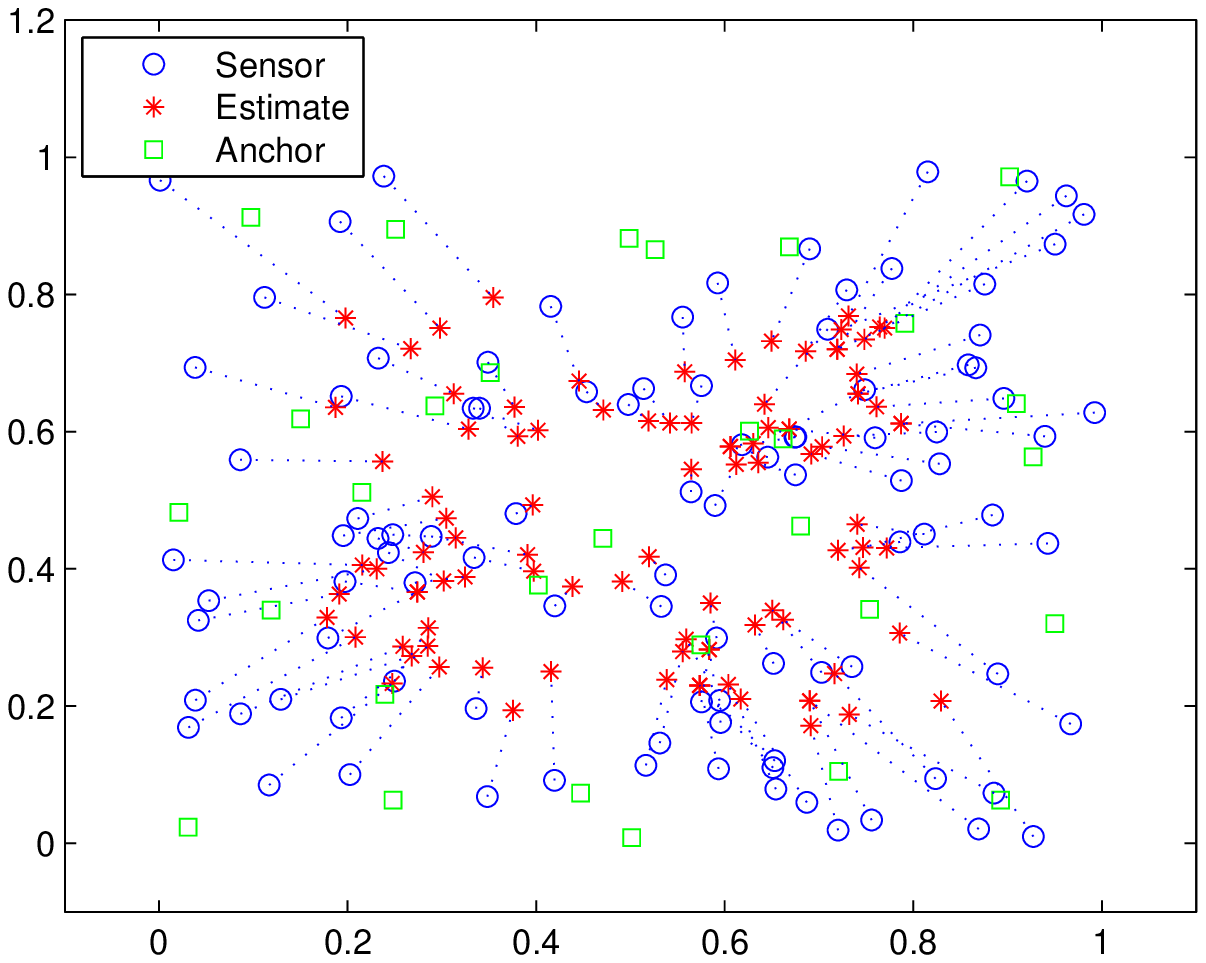}}
\caption{Location error with different transmission ranges}
\label{fig:first_placement} 
\end{figure}

Fig. 8 plots $e(NA)$ for both CA and RLA as the number of sensors varies from 50 to 80 with the same transmission range, at $DOI=0.1$. The number of anchors $NA=30$.

Fig. 9 plots $e(NA)$ for both CA and RLA as the number of sensors varies from 50 to 80 with different transmission ranges, at $DOI=0.1$. The number of anchors $NA=30$. Comparing Fig. 8 with Fig. 9 shows that the accuracy gain of RLA over CA is higher when the anchors have different ranges, than when they have the same range.

\begin{figure}
\centering
\includegraphics[width=3.5in]{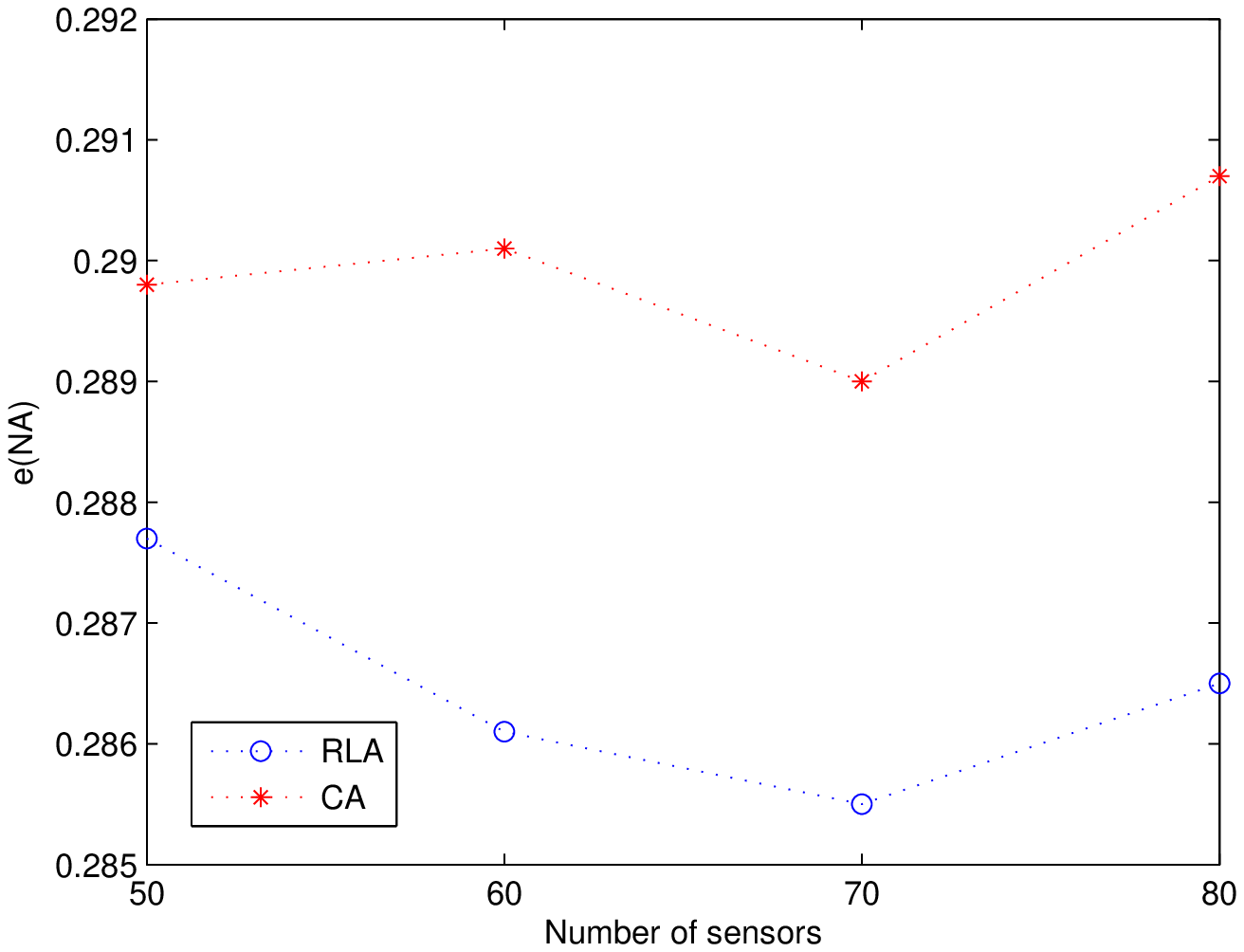}
\caption{The average localization error vs. the number of sensors (DOI=0.1 and the same transmission range)} \label{fig:sensor number and DOI}
\end{figure}

\begin{figure}[t]
\centering
\includegraphics[width=3.5in]{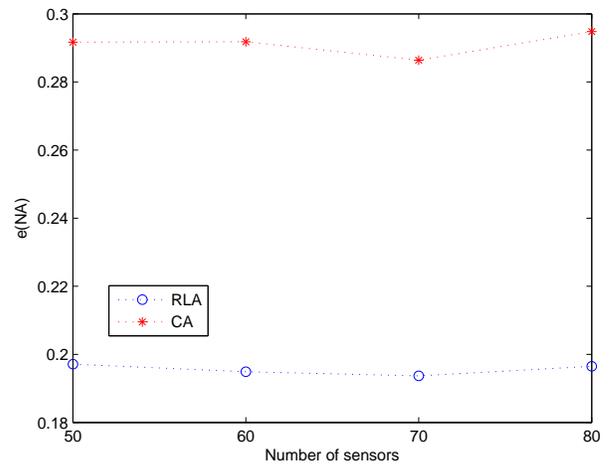}
\caption{The average localization error vs. the number of sensors (DOI=0.1 and different transmission ranges)} \label{fig:sensor number and DOI}
\end{figure}

\section{Conclusions}
Range-free localization, while not as accurate as range-based, has the principal advantage of simplicity, i.e., there is no requirement for special hardware to measure time-of-arrival or time-difference-of-arrival. This is important for WSN in which sensors are low cost units, and in some applications where knowing accurate sensor positions is not critical. While determining $\hat P$ from CA is simple, it is possible to improve on its accuracy with some additional computations. The RLA provides such an option and the simulation results in Section III show that there is approximately a 2 to 30\% gain in accuracy, depending on whether or not the anchors have identical ranges, and on the value of DOI.

\section*{Acknowledgement}
This work was supported by the CREST Advanced Integrated
Sensing Technology project of the Japan Science and Technology
Agency.

\end{document}